\documentclass{ws-procs9x6}

\begin{document}

\title{Physics results of the Pierre Auger Observatory}

\author{V. VAN ELEWYCK$^*$ \\
for the Pierre Auger Collaboration}

\address{Institut de Physique Nucl\'eaire d'Orsay, Universit\'e de Paris Sud  \& CNRS-IN2P3 \\
15, rue G. Clemenceau, 91406 Orsay Cedex, France\\
$^*$E-mail: vero@ipno.in2p3.fr}



\begin{abstract}

\end{abstract}


\bodymatter

\section{Introduction}
\label{sec:intro}

With three operational fluorescence sites out of four and more than 1000 active Cherenkov detectors on the ground at the time of writing, the Pierre Auger Observatory is nearing completion and has started accumulating data at a regularly increasing pace. In spite of the still small statistics available, a lot of progress has been made in the understanding and fine-tuning of the detector, which has resulted in the development of reliable analysis methods and of the release of its first scientific results concerning the main issues in ultra-high energy (UHE) cosmic ray physics. The spectrum of UHE cosmic rays observed by Auger is presented in \cite{paolo} as an illustration of the power of Auger's hybrid detection, and the present contribution will focus on the results obtained in the context of anisotropy searches and composition studies.   

\section{The arrival direction of UHECR: anisotropy studies with the Auger Observatory}
\label{sec:anis}

Anisotropies in  the flux of UHE cosmic rays may appear in different
energy ranges and angular scales, depending on the nature, distance
and extension of the source. Cosmic rays around an EeV are thought to be of galactic origin, and 
the region of the Galactic Center and
the Galactic Plane are key targets for anisotropy searches performed with Auger data. At higher energies one rather expects UHE cosmic rays to
come from extra-galactic sources; 
a search for directional excesses of cosmic rays could then 
reveal a correlation with some (un)known astrophysical objects or even
 exotic sources.

The anisotropy studies performed by  Auger are based on all surface detector (SD) events (plus some hybrids) with zenith angle $\theta < 60^\circ$ that pass the quality cut T5, which requires that
the station with the highest signal be surrounded by a hexagon of working stations,
ensuring a good reconstruction of the event. The energy of the events is determined using the constant intensity cut method
and calibrating the $S_{38}$ parameter to the energy obtained from the florescence detector (FD) as described in \cite{paolo}.

\subsection{Angular resolution and coverage maps}
To detect an excess of events coming from a particular region of the
sky, one has to compare the number of events observed in that region
with the corresponding coverage map, that is, the background number
of events expected from an isotropic flux of cosmic rays in the same
exposure conditions. This procedure requires accurate knowledge
of the detector properties, and in particular of its angular
accuracy and of the exposure dependance in time, energy and solid
angle for each point of the sky. For a more detailed discussion of these issues, we refer the reader to \cite{ICRCAA,cris}. 

The angular resolution $AR$ for the SD is defined as the angular radius
that would contain $68\%$ of the showers coming from a point source; it is
determined from the zenith ($\theta$) and azimuth
($\phi$) uncertainties obtained from the geometrical reconstruction on an event by event basis,
\begin{equation}
AR = 1.5\ \sqrt{\left[ \sigma^2 (\theta) + sin^2(\theta)\,\sigma^2 (\phi)\right]/2}
\end{equation}
 where $\sigma^2$ is for the variance. The $AR$ is driven by the accuracy on
 the measurement of the arrival time of the shower front in each station. The variance
 on the arrival time $T_1$ of the first particle is parameterized according to the
time variance model described in \cite{ICRCAA, cris}, which was
validated using data from the so-called "doublets" (pairs of tanks
separated by 11 m).

To build the coverage maps, one has to consider all possible
modulations and inhomogeneities in the exposure of different regions of the sky.
Besides the obvious effects due to the rotation of the Earth and the
limited field of view of the detector, other modulations are induced
by the continuously growing size of the array, by temporary failures
in some detectors, and by temperature and pressure variations (which
affect both the shower development in the atmosphere and the
response of the electronics). Two different techniques have been
used to estimate the SD coverage maps \cite{ICRCcoverage}:
\begin{itemize}
\item {\bf the semi-analytic method} consists in an analytical fit to the $\theta$
distribution of the events in the relevant energy range, convoluted
with an acceptance factor which accounts for the time evolution of
the detector (according to the trigger activity), assuming a uniform
response in azimuth (which is valid for showers up to $60^\circ $).
\item {\bf the shuffling technique} takes the average of many fake data sets
generated by shuffling the observed events in such a way that the
arrival times are exchanged and the azimuths are drawn uniformly.
This shuffling also 
preserves the $\theta$
distribution of the events. 
It might partially absorb an intrinsic large-scale
anisotropy present in the cosmic ray flux, but this drawback can be
avoided using independent shufflings in (day x hour).
\end{itemize}

The expected number of events in a given pixel of the sky is
obtained by integrating the coverage map in a given window, while
the signal is determined by applying the same filtering to the event
map. A significance map is then generated by comparing the signal in each
pixel respect to the expected background, according to the Li \& Ma
procedure \cite{LiMa}.

\subsection{Anisotropy studies around the galactic center}\label{subsec:GC}
The region of the Galactic Center (GC, located at
the equatorial coordinates  $(\alpha,
\delta)=(266.3^\circ ,-29.0^ \circ)$) and the Galactic Plane (GP) are 
 particularly attractive
targets for cosmic ray anisotropy studies around EeV energies. Two
cosmic ray experiments, AGASA and SUGAR, have already claimed
significant excesses in the flux of UHECR in that region. 
AGASA \cite{AGASACG} reported a $4.5\sigma$ excess of CR with
energies in the range $10^{18}-10^{18.4}$ eV in a $20^\circ$ radius
region centered at  $(\alpha, \delta) \simeq (280^\circ ,
-17^\circ)$ (it is worth noting however that the GC itself lies
outside of the AGASA field of view). Subsequent searches near this
region using old SUGAR data \cite{SUGAR} failed to confirm that
result but found a $2.9\sigma$ excess flux of CR in the energy range
$10^{17.9}-10^{18.5}$ eV in a $5.5^\circ$ window centered at
$(\alpha, \delta) \simeq (274^\circ , -22^\circ)$. Recent observations 
by HESS of a TeV $\gamma$
ray source in that region \cite{HESSCG} and of diffuse $\gamma$-ray emission from
the central 200 pc of the GP \cite{HESSGP} have provided additional hints towards
the presence of powerful CR accelerators in the Galaxy. In that
context, several models that predict a detectable flux of neutrons
in the EeV range (whose decay length is about the distance from the
GC to the Earth) have also been proposed.

\begin{figure}[t]
\begin{minipage}[t]{0.57\textwidth}
\mbox{}\\
\centerline{\includegraphics[width=\textwidth]{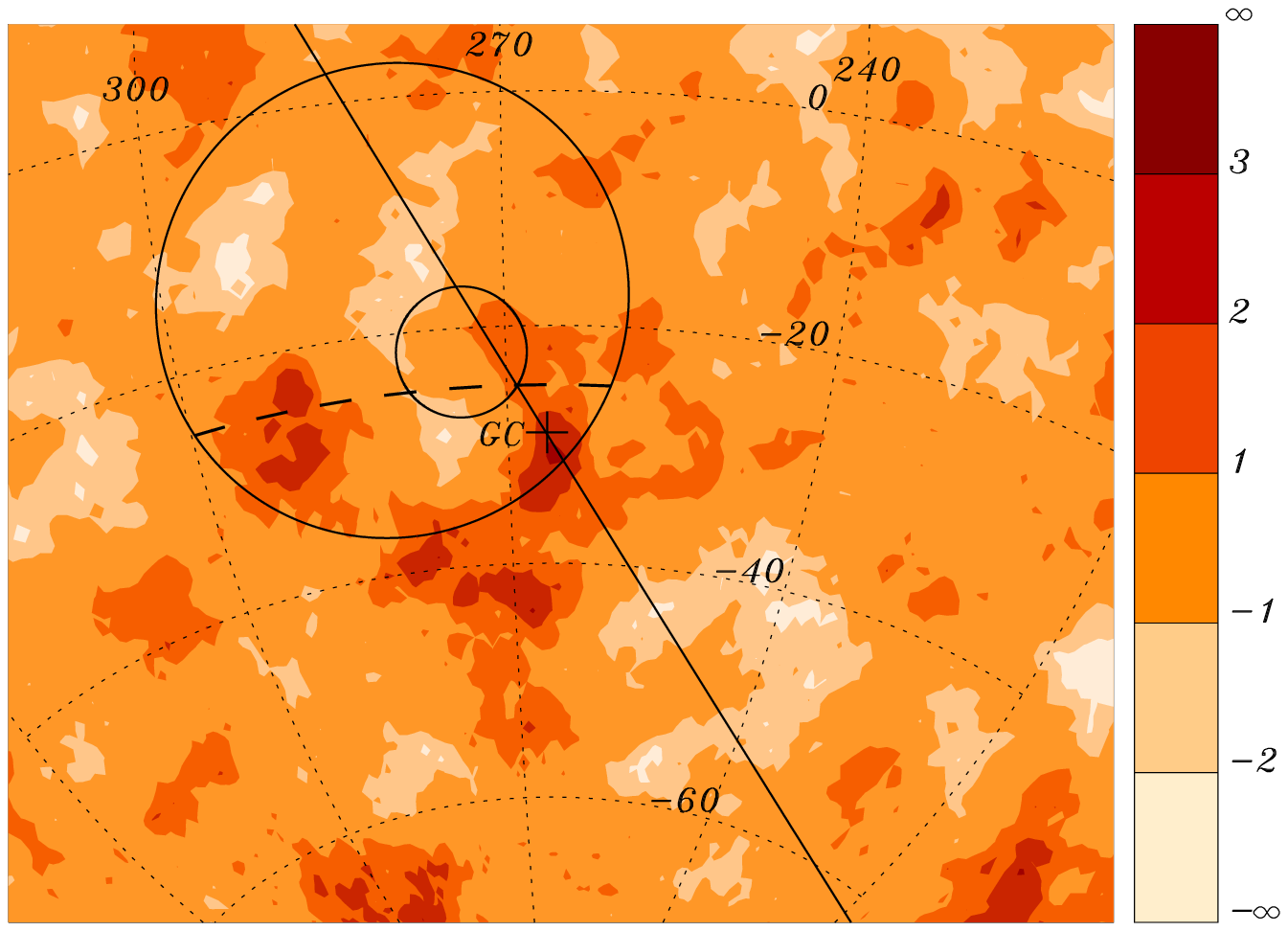}}
\end{minipage}
\begin{minipage}[t]{0.38\textwidth}
\mbox{}\\
\centerline{\includegraphics[width=0.9\textwidth,angle=270]{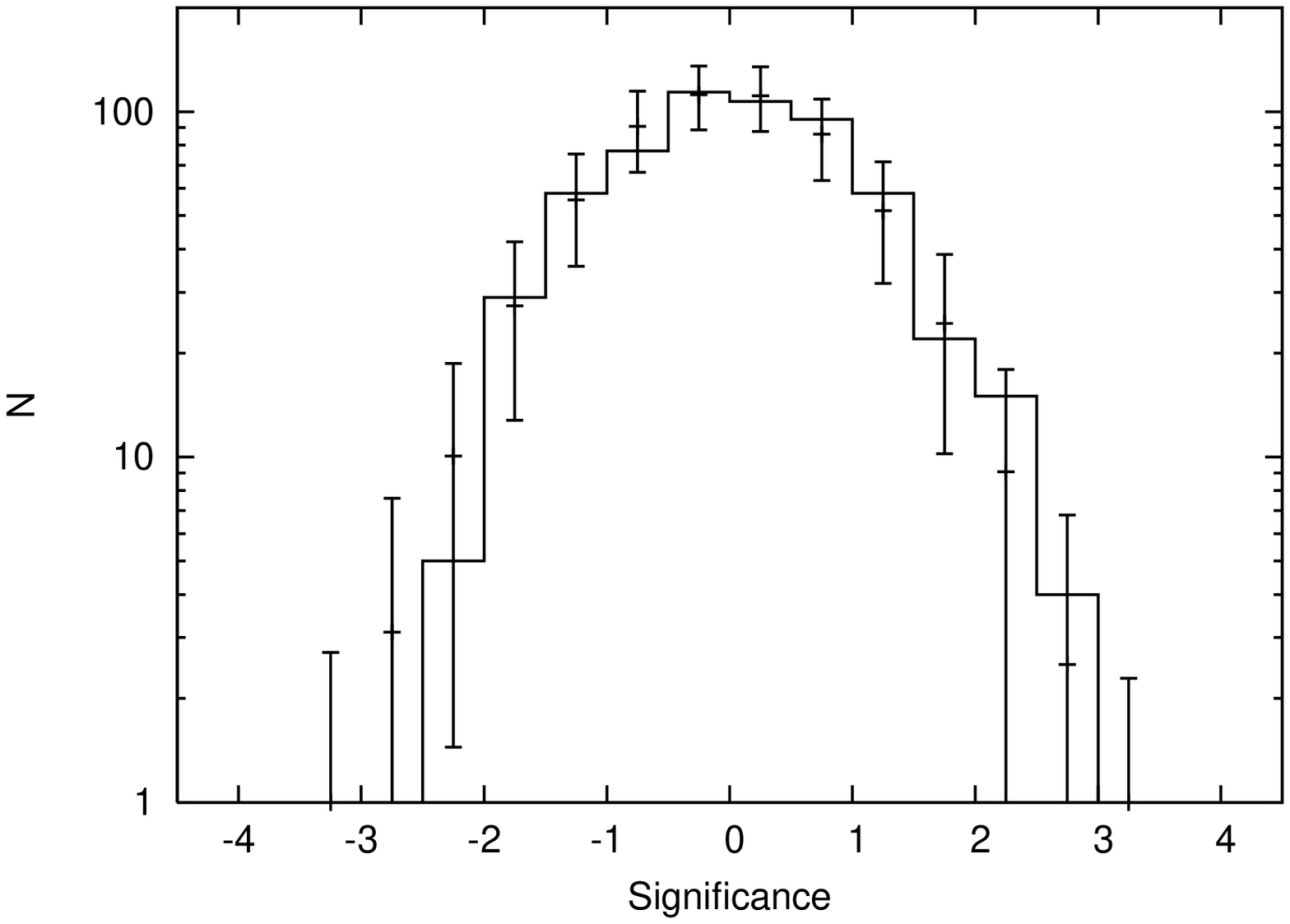}}
\end{minipage}
\caption{{\it Left: }significance map of CR overdensities in the region of the Galactic Center in the energy range $10^{17.9} - 10^{18.5}$ eV,  showing the Galactic Center (cross), the Galactic Plane (solid line), the regions of excesses of AGASA and SUGAR (circles), and the AGASA
field of view limit (dashed line). The event map was smoothed with a top-hat $5.5^\circ$ window.
{\it Right: } corresponding histogram of overdensities computed on a grid of $3^\circ$ spacing, compared to the average isotropic expectations points (with $2\sigma $ bars). (from \cite{AugerCG})}
\label{fig:GC}
\end{figure}

With the GC well in the field of view and an
angular resolution which is much better than previous CR
experiments, the Pierre Auger Observatory is well suited to look for
UHECR anisotropies coming from that region. A total of 79265 SD
events and 3934 hybrid events have been used, which corresponds to
the data collected between January 2004 and March 2006 satisfying
the T5 quality cut \cite{paolo} and with $\theta < 60^\circ, 10^{17.9}
\ \mathbf{eV} < E < 10^{18.5} \ \mathbf{eV}$; it represents respectively more than four and ten times the sample that AGASA and SUGAR used in this context.

Significance maps were
built using different filterings of the data to account for the
angular size of the excesses reported by AGASA and SUGAR in their
respective energy range. An example of such a map is shown in Fig. \ref{fig:GC} together with the corresponding overdensity distribution. Several tests were also performed with modified energy windows to account for a possible energy shift due to differences in the calibration of the experiments. In all cases, Auger data have been found compatible with isotropy, therefore not confirming the results from previous experiments. Even in the worst case of a source emitting nucleons and
embedded in a background made of heavier nuclei, to which Auger is more sensitive in the relevant energy
range, a significant excess ($~ 5.2 \sigma$) would be expected, in contradiction with current observations. Details of the analysis can be found in \cite{AugerCG}.

Data from the Auger Observatory were also used to search for a point
source in the direction of the GC itself at the scale of Auger's own 
angular resolution. In the energy range $10^{17.9}$ --
$10^{18.5}$ eV, and applying a $1.5^\circ$ Gaussian filter to
account for the pointing accuracy of the SD, we obtain 53.8 observed
events against 45.8 expected. This allows to put a $95\%$ C.L. upper
bound on the number of events coming from the source of $n_s^{95} =
18.5$. Assuming that, in the energy range considered, both the
source and the bulk CR spectrum have similar spectral indexes and
that the emitted CR are proton-like, and taking a differential spectrum $\Phi_{CR}(E) \simeq \xi\ 30\
(E/EeV)^{-3}\  EeV^{-1} km^{-2}yr^{-1}sr^{-1},$ where $\xi$
parameterizes the uncertainties on the flux normalization, a 95\% C.L. upper
bound of
\begin{equation}
\Phi_s^{95} \leq \xi \ 0.08\ km^{-2} yr^{-1}
\end{equation}
can be set on the source flux. This bound could however be about $30\%$ higher if the
CR composition at EeV were heavy, {\it i.e.} close to Iron.

Finally, a scan for correlations of CR arrival directions with the Galactic 
Plane and Super-Galactic Plane have also been made in two different windows of 
energy (1 EeV $<$ E $<$ 5  EeV and  E $>$ 5  EeV), yielding again negative 
results (although with a smaller dataset) \cite{ICRCGC}. 

\subsection{Other searches for localized excesses in the Auger sky maps}
The Auger data have also been used to perform both blind searches and  prescripted searches for localized excesses in other parts of the sky.

In the case of blind searches, the distribution of significances is compared to those obtained from a large number of Monte Carlo isotropic simulations. Such searches were performed both for a $5^{\circ}$ and a $15^{\circ}$ angular scale and in two separate energy ranges, $1 \mathbf{EeV}\leq E \leq 5 \mathbf{EeV}$ and  $  E \geq 5 \mathbf{EeV}$; all of them turned out to be compatible with isotropy \cite{ICRCbenoit}.

The Auger Collaboration had also released  a list of prescribed targets with definite angular and energy windows \cite{ICRColdpres}, with the associated significance probability level to attain in order to claim a positive signal. The prescription targets range from the Galactic Center to some nearby violent extragalactic objects; none of them has turned out to lead to a positive detection. As more data is streaming in, the catalogue of candidate targets that will be studied is expected to increase in the future.

\section{The nature of UHECR: composition studies with the Auger Observatory}

Thanks to its hybrid capabilities, the Auger Observatory can extract complementary information on the shower development parameters, that are ultimately related to the nature of the primary cosmic rays. If the discrimination between different types of nuclei is complicated by the uncertainties in the hadronic models governing the interactions of the particles in the shower at such high energies, several methods have already been proposed for the identification of photons and neutrinos, and are currently applied to the data of the Auger Observatory. The presence and the amount of photons and neutrinos at such high energies would constitute a crucial probe for many exotic models of UHE cosmic ray production and could help locate candidate sources as they travel undeflected by the intergalactic magnetic fields. 

\subsection{Upper limit on the UHE photon flux}
\label{sub:photon}
Unlike protons and nuclei, the development of photon showers are driven by electromagnetic (EM) interactions and do not suffer much from the uncertainties in hadronic interactions. Photon showers are expected to contain fewer and less energetic secondary muons, as a result of the smallness of the photon radiation length respect to its mean free path for photo-nuclear interactions and direct muon production. Their development is also delayed due to the small multiplicity in EM interactions and
 to the LPM effect \cite{LPM}, 
which reduces the bremsstrahlung and pair production cross-sections at energies above 10 EeV. These considerations allowed several ground array experiments to set upper limits on the flux of UHE photons on basis of studies of the rate of vertical to inclined showers (in Haverah Park experiment \cite{HPphoton}) and of the muon content of the showers at ground (in AGASA \cite{AGASAphoton}).

\begin{figure}[t]
\begin{minipage}[t]{0.46\textwidth}
\mbox{}\\
\centerline{\includegraphics[width=\textwidth]{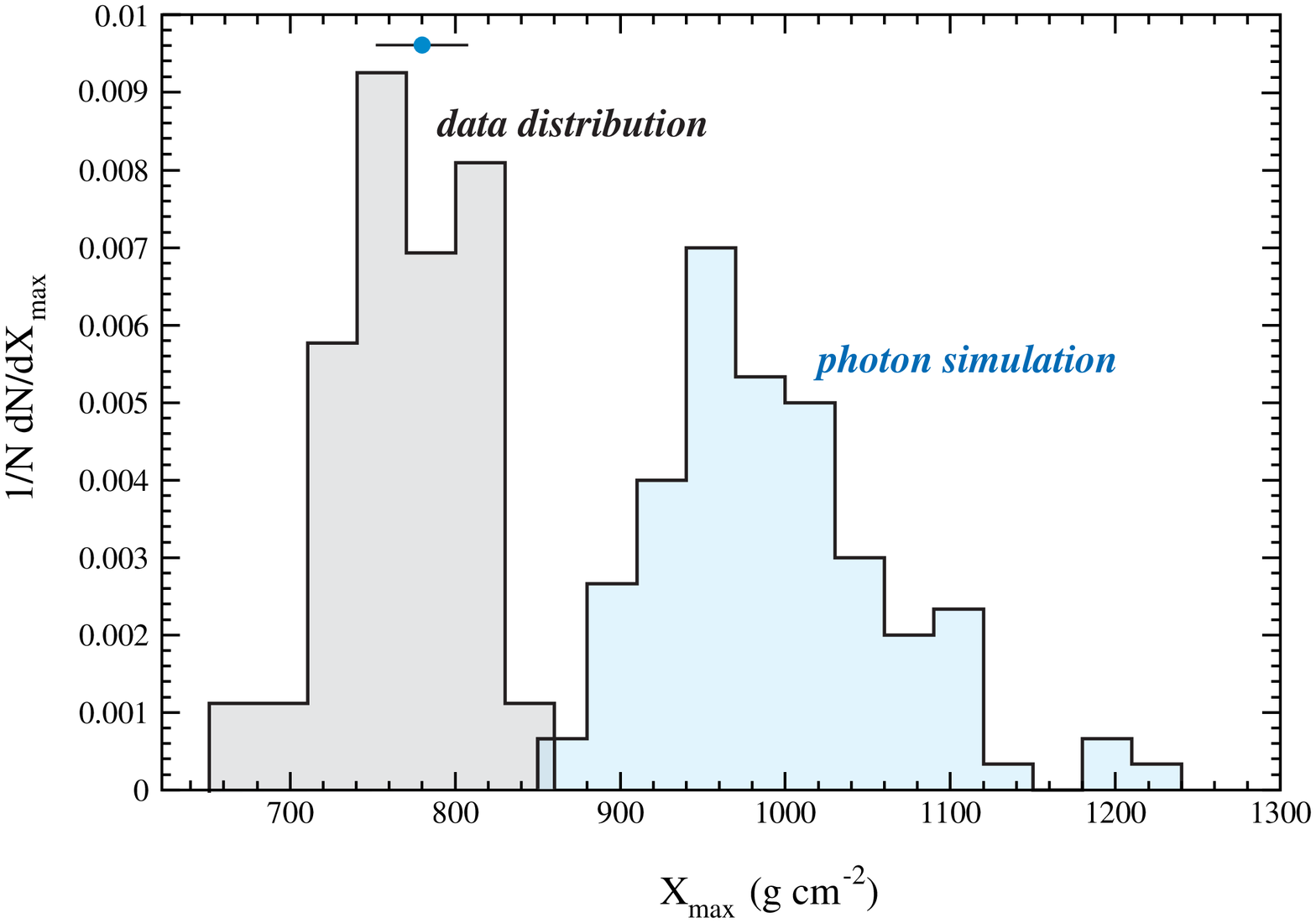}}
 \caption[Xmax vs. energy] {\raggedright $X_{max}$ distribution of simulated and real candidate photon events. The blue point is the $X_{max}$ value and uncertainty for one event from the data.}
\label{fig:xmaxcomp}
\end{minipage}
\hfill
\begin{minipage}[t]{0.48\textwidth}
\mbox{}\\
\centerline{\includegraphics[width=\textwidth]{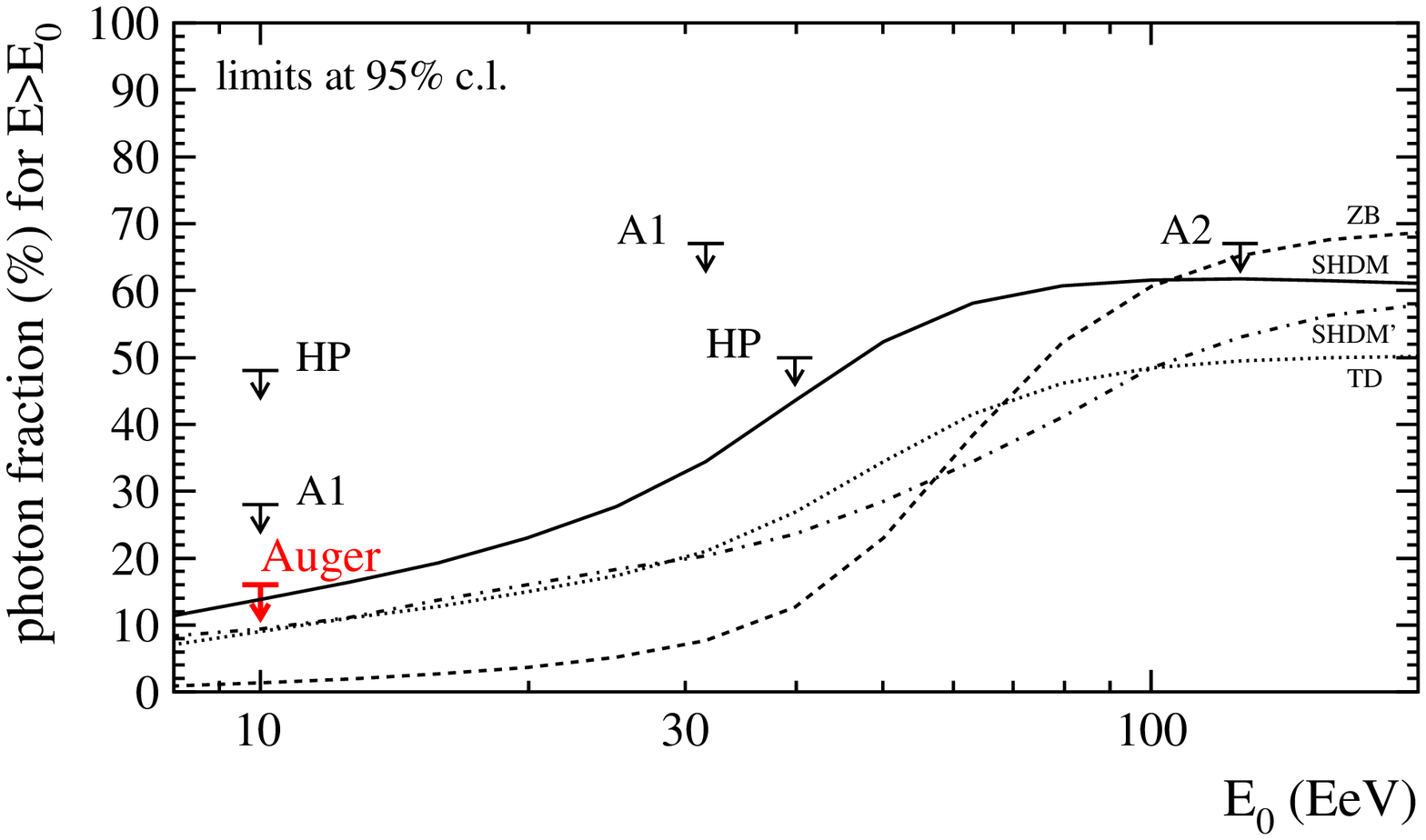}}
  \caption[photon fraction upper limits]{\raggedright  95\% C.L upper limit on the photon fraction in UHE cosmic rays obtained by Auger, compared to the results of Haverah Park \cite{HPphoton}, HP, and AGASA \cite{AGASAphoton}, A1 and A2.}
\label{fig:upperlimit}
\end{minipage}
\end{figure}

Taking profit of its hybrid design, Auger has set up a different method to identify photon primaries in the flux of UHECR. It is based on the direct observation of the longitudinal profile of the shower development in the atmosphere by the FD, and uses as a discriminating variable the atmospheric depth of the shower maximum, $X_{max}$ (the estimated average difference in $X_{max}$ between photons and hadrons is about $200 \mathrm{gr/cm}^2$).

The data set used for this analysis corresponds to the hybrid events ({\it i.e.} those observed by one or more FD telescope and by at least one SD station, which  ensures a better angular accuracy  and smaller uncertainty in the reconstruction of $X_{max}$) with a reconstructed energy $E > 10^{19}$ eV, registered between January 2004 and February 2006. During that period two of the four Auger eyes were active (for a total of 12 FD telescopes) and the number of deployed SD stations grew from 150 to 950. 

A series of cuts were applied to the data that guarantee the quality of the hybrid geometry and of the fit to the shower longitudinal profile, which takes into account the local amtospheric conditions (see the detail in \cite{Augerphoton}). One important  condition is to have the {$X_{max}$} of the shower inside the field of view of the telescopes. To minimize the bias that this condition introduces against photon primaries in the detector acceptance, additional energy-dependant cuts are applied both on the  zenith angle and the maximum distance of telescope to shower impact point in order to eliminate nearly-vertical and distant events. 

For each of the 29 events that survived all the cuts, 100 photon showers were simulated in the same energy and arrival direction conditions and the resulting expected distribution of {$X_{max}$} was compared to the observed $X_{max}$ of the event. An example is shown in Fig. \ref{fig:xmaxcomp}, together with the distribution of the $X_{max}$ from the whole selected dataset. For all 29 events, the observed $X_{max}$ is well below the average value expected for photons.
Taking systematic uncertainties on the $X_{max}$ determination and the photon shower simulations into account, the available statistics allows to put an upper limit on the photon fraction of 16\% at 95\% C.L, which is shown in Fig. \ref{fig:upperlimit} together with previous results and some predictions from non-accelerator models.

\subsection{Inclined air showers and the detection of neutrinos}

\begin{figure}[t!]
\centerline{\includegraphics[width=0.54\textwidth]{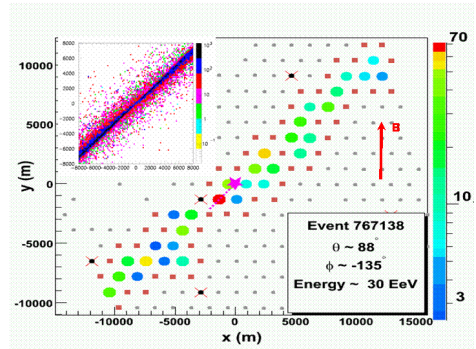}}
\caption{Example of a near horizontal air shower as seen by the SD; the shower triggered 31 tanks and extends on about 30 km at ground. In the upper left corner, the best fitting simulated muon map corresponding to the reconstructed zenith and azimuth angles.} 
\label{fig:HAS}
\end{figure}

The use of Cherenkov water tanks for the SD allows the Pierre Auger Observatory to detect showers with zenith angles up to $90^\circ$ (and even more)\cite{HAS}. The range of inclined showers, $60^\circ \leq \theta \leq 90^\circ$, contributes half the total solid angle of the detector and about 25\% of its geometrical acceptance, thereby significantly increasing the field of view of the detector and the SD statistics. Such events are indeed seen by Auger both in the SD and the FD; some of them may be quite spectacular, with very extended footprints involving tens of tanks, as illustrated in Fig. \ref{fig:HAS}. Dedicated selection procedures and reconstruction methods are being developed in Auger to deal with the distinctive features of those showers.  

\begin{figure}[t]
\begin{minipage}[t]{0.40\textwidth}
\mbox{}\\
\centerline{\includegraphics[width=\textwidth]{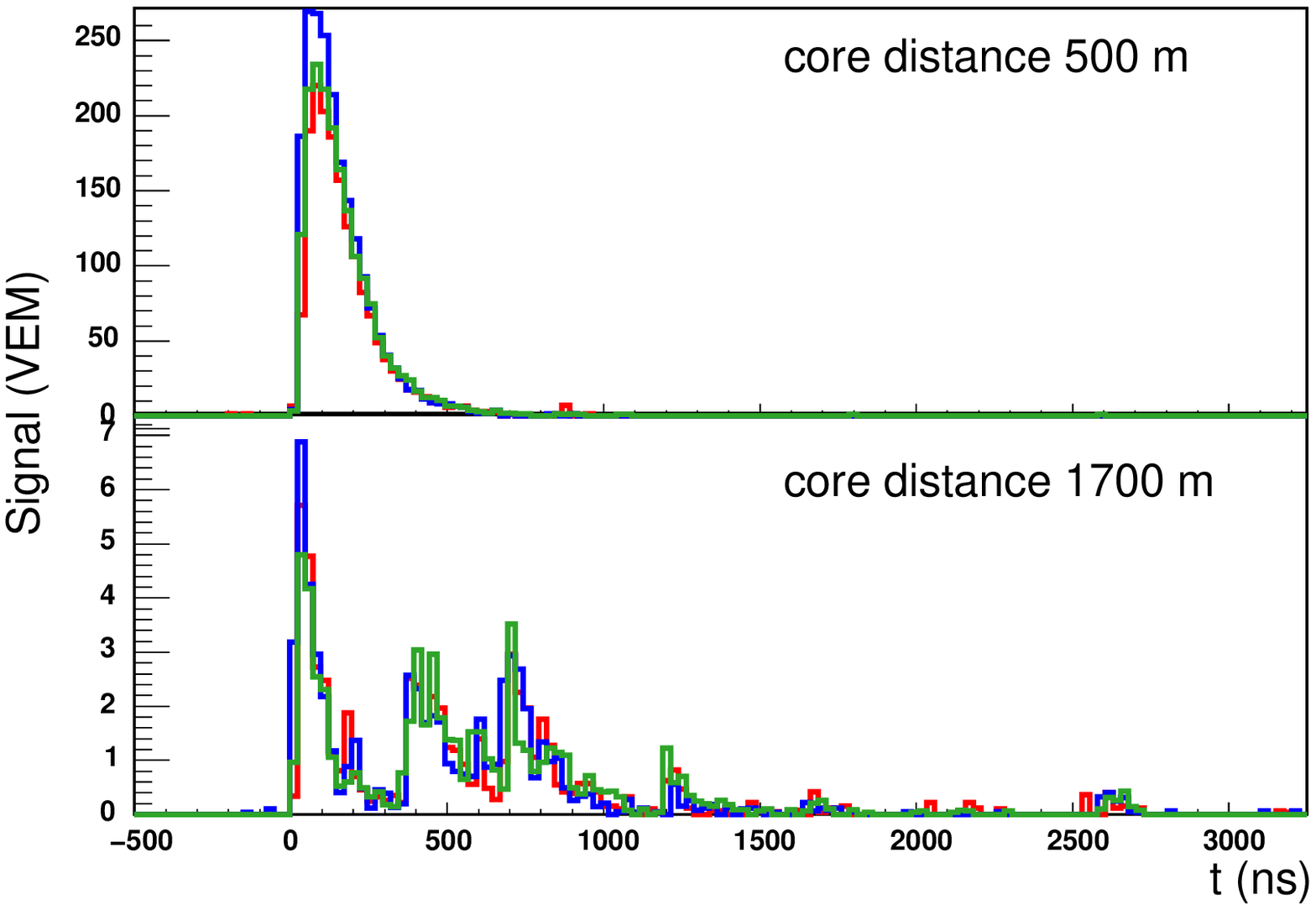}}
\end{minipage}
\hfill
\begin{minipage}[t]{0.38\textwidth}
\mbox{}\\
\centerline{\includegraphics[width=\textwidth]{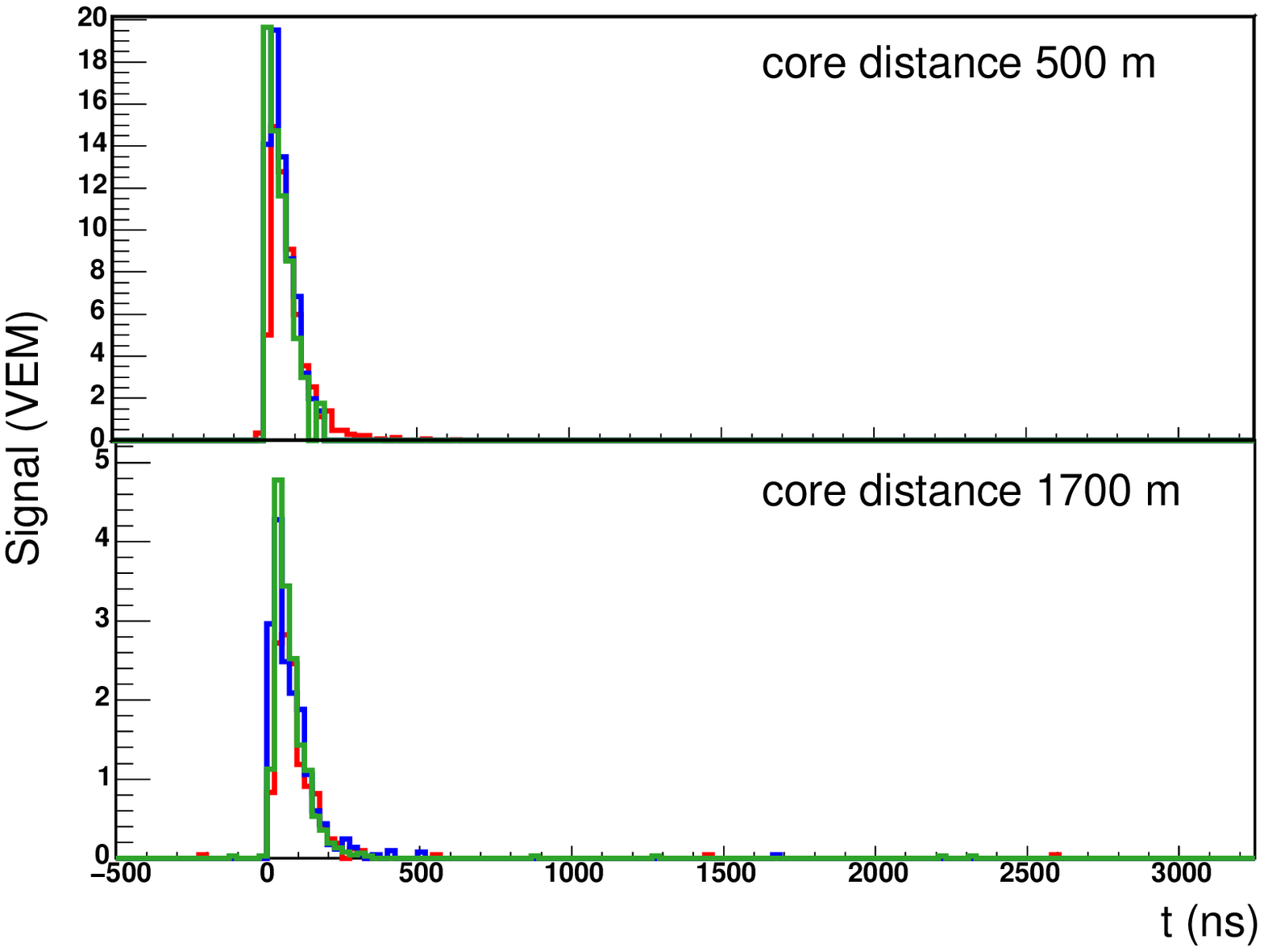}}
\end{minipage}
 \caption[FADC] {FADC traces of a young ({\it left}) and old ({\it right}) shower. The signal from a young shower gets smaller and more extended as the distance to the core increases, while old showers have short traces at all distances.}
\label{fig:FADC}
\end{figure}

The distance between the first interaction point (normally in the first few 100 g  $\mathrm{cm}^{-2}$) and the detector position is much  larger than in the vertical case, the atmospheric depth ranging from 1740 g $\mathrm{cm}^{-2}$ at $60^\circ$ till 31 000 g  $\mathrm{cm}^{-2}$  at $90^\circ$). 
As a result, the EM  component of the shower dies out long before reaching the ground, and the only particles recorded in the SD are energetic muons (typically of 10-1000 GeV) accompanied by an EM halo which is constantly regenerated by muon decay, brehmsstrahlung and pair production. Those muons arrive at ground in a thin front with small curvature, resulting in short FADC pulses in the tanks, as shown in Fig.\ref{fig:FADC} (right). Their trajectories are long enough to be affected by the geomagnetic field, which leads to a separation between positive and negative muons and a further elongation of the projected footprint on the ground.

The reconstruction of inclined showers is based on the search for the best fit to the pattern of signals at ground performed with averaged maps of muon densities obtained from simulations. The relation between the muon density and the energy depends on the nature of the primary cosmic ray, and is established on basis of Monte Carlo simulations which suffer from hadronic interactions uncertainties at high energies. In this context, hybrid inclined events reconstructed by both the SD and the FD will play an important r\^{o}le in primary composition studies, since they allow independant measurements of the EM and muonic components of the shower \cite{ave}.

Inclined showers also constitute the bulk of events from which a signal of UHE cosmic neutrino could be extracted. Due to their small cross-section, neutrinos can penetrate deeply in the atmosphere and initiate showers at all possible depths, unlike nuclei or photons. In particular, showers originating less than $\approx 2000$ g $\mathrm{cm}^{-2}$ away from the detector will reach it before their EM component attenuates completely. Selection criteria will thus require the presence of signals corresponding to a young shower, and in particular of stations with extended traces that reflect the large curvature of the shower front and the presence of an EM component (see Fig. \ref{fig:FADC}). 

Up-going tau neutrinos that skim the Earth just below the horizon could also be detected as they are likely to interact in the ground and produce a tau which may emerge from Earth and initiate an observable air shower, provided it decays close enough to the SD.  Preliminary studies provided a proof of principle for the detection of such neutrinos in the energy range $10^{17}--10^{19}$ eV \cite{nu} and, although a careful study of systematic uncertainties is necessary to infer with a reasonable precision the energy of the incident $\nu_\tau$ primary, this method seems the most promising in terms of acceptance, which is a crucial matter when dealing with event rates as small as $\sim$ 1 per year. Studies are currently ongoing both in the down-going and upgoing ranges to define and optimize the selection criteria, and the search for UHE neutrinos in the Auger data has started.
  
\section{Conclusions}

The Southern Auger Observatory, expected to be complete in 2007, has delivered its first science results on the UHE cosmic ray spectrum, anisotropy searches and composition studies. In particular, the region of the Galactic Center has been studied with a precision never attained before, yielding no hint of anisotropies. The absence of evidence for a point-source near the GC excludes several scenarios of neutron sources recently proposed. The upper limit on the photon fraction above 10 EeV, derived for the first time from a direct observation of the shower maximum, confirms and improves previous limits from ground arrays. Finally, the inclined shower data sample will soon contribute to enlarge the field of view of the detector and increase its statistics; it might also reveal the first cosmic neutrino ever observed at ultra-high energies.   
\section*{Acknowledgements}

Many warm thanks are due to the directors and organizing staff of the School, profs. M. Shapiro, T. Stanev, J. Wefel and A. Smith, 
for generating a lively and inspiring scientific atmosphere in Erice. I am grateful to prof. J. Cronin for giving me the opportunity to present the Auger results to such a rewarding audience. This work was supported by the European Community $\mathrm{6^{th}}$ F.P. through the Marie Curie Fellowship MEIF-CT-2005 025057.


\begin{thebibliography}{9}

\bibitem{paolo} P. Privitera, {\it The Auger Observatory}, these Proceedings.

\bibitem{ICRCAA}  C. Bonifazi [Pierre Auger Collaboration], Proc. 29th ICRC {\bf 7} (2005) 17.

\bibitem{cris}A.~Letessier-Selvon  [Pierre Auger Collaboration],
  arXiv:astro-ph/0610160.

\bibitem{ICRCcoverage}  J.-Ch. Hamilton [Pierre Auger Collaboration], Proc. 29th ICRC {\bf 7} (2005) 63.

\bibitem{LiMa} T.-P. Li and Y.-Q. Ma, Astrophys. J {\bf 272} (1983) 317.

\bibitem{AGASACG} N.~Hayashida {\it et al.}  [AGASA Collaboration],
  Astropart.\ Phys.\  {\bf 10} (1999) 303
; M. Teshima {\it et al.} [AGASA Collaboration], in Proc. 27th ICRC {\bf 1} (2001) 337.


\bibitem{SUGAR} J.~A.~Bellido {\it et al.}, 
  Astropart.\ Phys.\  {\bf 15} (2001) 167.

\bibitem{HESSCG} F.~Aharonian {\it et al.}  [HESS Collaboration],
  Astron.\ Astrophys.\  {\bf 425} (2004) L13.

\bibitem{HESSGP} F.~Aharonian {\it et al.}  [HESS Collaboration],
  Nature {\bf 439} (2006) 695.


\bibitem{AugerCG} M.~Aglietta {\it et al.}  [Pierre Auger Collaboration],
Astropart. Phys., in press [arXiv:astro-ph/0607382].

\bibitem{ICRCGC}   A. Letessier-Selvon [Pierre Auger Collaboration] Proc.29th ICRC {\bf 7}(2005) 67.

\bibitem{ICRCbenoit}   B. Revenu [Pierre Auger Collaboration], Proc. 29th ICRC {\bf 7} (2005) 75.

\bibitem{ICRColdpres} R. Clay [Pierre Auger Collaboration], Proc. 28th ICRC {\bf 1} (2003), 421.

\bibitem{LPM} L. D. Landau, I. Ya. Pomeranchuk Dokl. Akad. Nausk. SSSR {\bf 92} (1953), 535 \& 735; A. B. Migdal, Phys. Rev. {\bf 103} (1956), 1811.

\bibitem{HPphoton} M. Ave {\it et al.}, Phys. Rev. Lett. {\bf 85} (2000), 2244; Phys. Rev. {\bf D65} (2002) 063007.

\bibitem{AGASAphoton} K. Shinozaki {\it et al.}, Astrophys. J. {\bf 571} (2002), L117; M. Risse {\it et al.}, Phys. Rev. Lett. {\bf 95} (2005),171102.

\bibitem{Augerphoton} J. Abraham {\it et al.}   [Pierre Auger Collaboration],
Astropart. Phys., in press [arXiv:astro-ph/0606619]. 

\bibitem{HAS} L. Nellen [Pierre Auger Collaboration], Proc. 29th ICRC {\bf 7} (2005), 183; V. Van Elewyck [Pierre Auger Collaboration], AIP Conf. Proc. {\bf 819} (2006), 187.

\bibitem{ave} M. Ave {\it et al.}, Proc. 28th ICRC {\bf 1} (2003), 563.

\bibitem{nu} K.S. Capelle {\it et al.}, Astropart. Phys. {\bf 8} (1998), 321;
X. Bertou {\it et al.}, Astropart. Phys. {\bf 17} (2002), 183. 
\end{thebibliography}
\end{document}